# Dual-Layer Video Encryption using RSA Algorithm

Aman Chadha
Electrical and Computer Engineering
University of Wisconsin-Madison

Sushmit Mallik
Electrical and Computer Engineering
NC State University

Ankit Chadha
Electrical and Computer Engineering
University of Minnesota

Ravdeep Johar
Computer Sciences
Rochester Institute of Technology

M. Mani Roja
Electronics and Telecommunication Engineering
Thadomal Shahani Engineering College (Univ of Mumbai)

## ABSTRACT
This paper proposes a video encryption algorithm using RSA and Pseudo Noise (PN) sequence, aimed at applications requiring sensitive video information transfers. The system is primarily designed to work with files encoded using the Audio Video Interleaved (AVI) codec, although it can be easily ported for use with Moving Picture Experts Group (MPEG) encoded files. The audio and video components of the source separately undergo two layers of encryption to ensure a reasonable level of security. Encryption of the video component involves applying the RSA algorithm followed by the PN-based encryption. Similarly, the audio component is first encrypted using PN and further subjected to encryption using the Discrete Cosine Transform. Combining these techniques, an efficient system, invulnerable to security breaches and attacks with favorable values of parameters such as encryption/decryption speed, encryption/decryption ratio and visual degradation; has been put forth. For applications requiring encryption of sensitive data wherein stringent security requirements are of prime concern, the system is found to yield negligible similarities in visual perception between the original and the encrypted video sequence. For applications wherein visual similarity is not of major concern, we limit the encryption task to a single level of encryption which is accomplished by using RSA, thereby quickening the encryption process. Although some similarity between the original and encrypted video is observed in this case, it is not enough to comprehend the happenings in the video.

## 1. INTRODUCTION
With multimedia becoming the norm for exchanging information in today's world, the security of commercial multimedia applications has assumed critical importance. For instance, enterprises with distributed locations having their business meetings via video conferencing, is now a commonplace. Having an intruder intercept the path of data-transmission and thereby gain access to the information being transferred can lead to horrendous situations especially in scenarios wherein sensitive data is being transferred.

Another related domain is the video-on-demand application wherein certain privileged users are granted access to receive the benefits of the service. To ascertain that the signal is not intercepted on its transmission path and hence prevent the misuse of the service, encryption can be used.

Such applications need stringent encryption algorithms for which the incurred expenses for cracking the encryption in terms of the cost should be more than the legal access to the service. This is to deter the misuse of the service. In addition, the time needed to crack the code should be significant to ensure a sufficient level of security.

Processing of large video files involves a huge volume of data. The codec, storage systems and network need high levels of resource utilization, i.e., processor time. Complex algorithms for encryption will only aggravate the problem and increase latency. Thus, the algorithm needed needs to be both secure as well as fast.

In this paper, we propose a system using the RSA algorithm. The RSA algorithm involves three steps: key generation, encryption and decryption.

Fig. 1 describes the process of encrypting and decrypting the video using two layers of encryption.

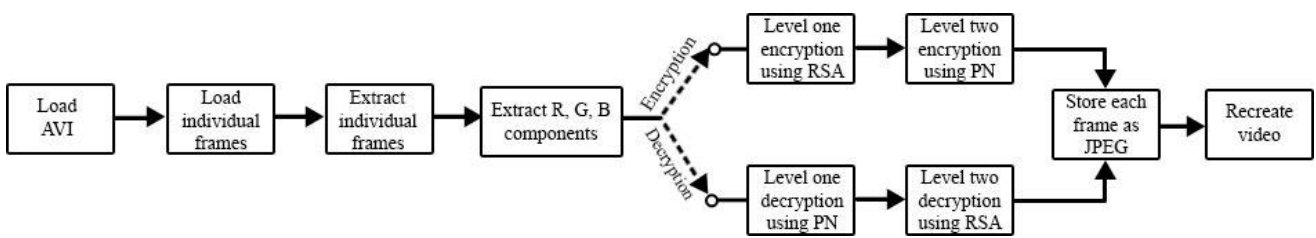

**Fig. 1 Process of Encryption and Decryption**

## 2. REVIEW OF LITERATURE
Cryptography is the art and science of protecting information from intruders to use it for malign purposes. It mainly comprises of Encryption and Decryption. Encryption deals with scrambling the content of a message to make it unreadable or undecipherable for any non-authorized personnel. The reverse of data encryption is Decryption which will reproduce the original data. The first known use of cryptography dates back to the ancient Egyptian civilization. Ever since then, cryptography has played an integral part in how we go about communication. It gained prominence during the Second World War when the allied forces gained an upper hand after they were able to break the German cipher machine, Enigma [1]. These days, it is recognized as one of the major components for providing information security, controlling access to resources and financial transactions. The original data to be transmitted is called Plain Text which is readable by a person or computer. When it is encrypted, it is known as Cipher Text. A system which provides encrypting and decryption is known as a cryptosystem. The level of security of an encryption algorithm is given by the size of its key space [3]. The larger the key space, the more complicated the encryption algorithm is. As a result it takes a considerably longer time to crack the code as compared to an encryption algorithm with a smaller key space.

Cryptography keys are usually classified as Symmetric and Asymmetric algorithms. In Symmetric Key Algorithms, the sender and receiver use the same keys for encryption as well as decryption. Symmetric Key Encryption is also known as secret key as both, the sender and receiver have to keep the key protected [4]-[5]. The level of security entirely depends on how well the sender and the receiver keep the key protected. If the unauthorized person is able to get the key, he can easily decipher the encryption using it. This is the major limitation of the Symmetric Key Algorithm. However, it is less computationally intensive and hard to break if a large key space is used.

Popular symmetric key algorithms are Data Encryption Standard (DES), Triple DES and Advance Encryption. DES is an example of block cipher, which operates on 64 bits at a time, with 64 bits as input key [8]. Every 8th byte is in the input key is parity check so, effectively the key space is reduced to 56 bits [6]-[8].

In 2001, the Rijndael cryptosystem was selected as the Advanced Encryption Standard (AES) [4], [8]. It operates on 128-bit blocks, arranged as $4 \times 4$ matrices with 8-bit entries. The algorithm can have variable block length and key length [8].

In 1976, the Asymmetric key algorithm, also called the Public Key Cryptography was first developed [8]-[9]. This enabled the sender and receiver to communicate over a non - secure interface because two different keys are used by the sender and the receiver instead of sharing a single key and making it vulnerable to external attacks. These keys are known as the public key and the private key. The private key is only known to the authorized user. The data is encrypted by one key and decrypted using the other [4], [8]. Both the keys are mathematically linked but it is possible to derive the private key from the public key. So, access to the private key should be protected as it is meant only for the authorized user [4]-[5]. The Rivest Shamir Adelman (RSA) algorithm is the most popular symmetric key algorithm. The RSA algorithm was developed by Ron Rivest, Adi Shamir and Len Adelman at MIT [25] in 1977. It is based on the idea of prime factorization of integers.

Comparatively, symmetric key encryption is much faster than asymmetric techniques [8]. In symmetric key algorithms, the security is dependent on the length of the key, unlike asymmetric algorithms. In symmetric key encryption, a secure mechanism is required to deliver keys properly while asymmetric keys result in better key distribution. Symmetric key provides confidentiality but not authenticity as the secret key is shared [8]. Asymmetric keys on the other hand, provide both authentication and confidentiality [8].

Next, we talk about the various video encryption schemes. In today's modern day where we depend on video transmission for news and media, protecting digital video from attacks during transmission is of prime importance. Due to the large size of digital videos, they are usually compressed and then transmitted using formats like MPEG [10] or H.264/AVC [11]. As such, the encryption formats also work in a compressed domain [8]. Many video encryption techniques have been proposed which make an attempt to optimize the encryption process with respect to the encryption speed and display process.

The Naïve Algorithm is the simplest way to encrypt every byte in the whole MPEG stream using standard encryption schemes like DES or AES. The idea of the Naïve algorithm is to treat the MPEG bit-stream as text data and does not use any special structure [12]-[14]. It ensures the security as no known can effectively break the AES or the triple DES encryption if a sufficiently large key space is used [8]. However, the Naïve algorithm is not an efficient solution for very large video content as it becomes very slow, especially while using the triple DES encryption. The resultant delays increase the encryption overload and thus are not so favorable for real time video transmission [8].

The Pure Permutation Algorithm scrambles the bytes within the frame of a MPEG stream by permutation. However, it has been demonstrated [15] that the pure permutation algorithm is vulnerable to known plaintext attack and therefore should be used cautiously [8]. By comparing the cipher text with known frames, the attacker can figure out the secret permutation and the frames can be deciphered. The basic steps are listed below [8]:

(1) A list of 64 permutations is generated.
(2) The splitting procedure is undertaken. It is assumed that the DC coefficient is denoted by 8 digit binary numbers, $d7, d6, d5, d4, d3, d2, d1, d0$. Then it is split into two numbers $d7\ d6\ d5\ d4$ and $d3\ d2\ d1\ d0$. The number $d7\ d6\ d5\ d4$ is assigned as DC coefficient and the number $d3\ d2\ d1\ d0$ is assigned as AC coefficient. The value of DC coefficient should be much larger than AC coefficient.
(3) Random permutation is applied to the split block.

The encryption and decryption add very little overhead to the video compression and decompression processes. However, it reduces the video compression rate as random permutation distorts the probability distribution of the Discrete Cosine Transform (DCT) and makes the Huffman table used less than optimal [8]. L. Qiao and Nahrstedt in 1998 introduced two attacks, the cipher text only attack and a known-plaintext attack.

Quiao and Nahresdt introduced a new algorithm called Video Encryption Algorithm (VEA) [17]. It is based on the statistical properties of the MPEG video standard and symmetric key algorithm which reduces the amount of data encrypted [8]. VEA divides the input video stream into two chunks (a1, a2, a3, a4, a5, a2n-1, a2n). These chunks are further divided in to data segments into even list (a2, a4, a6…a2n) and odd list (a1, a3, a5..a2n-1). After this, an encryption key is applied to the even list, E (a2, a4, a6…a2n), where E denotes the Encryption function used. The resultant encoded list is XORed with the odd list and the concatenated result is the final cipher text. As a result, the VEA is protected from known-only plain text attack as each frame will have a different key [8].

Four new algorithms were introduced by Bhargava, Shi and Wang [18]-[19]. The algorithms are Algorithm 1, Algorithm 2 (VEA), Algorithm 3 (MVEA), and Algorithm 4 (RVEA). Algorithm 1 uses the permutation of the Huffman code words in I-frames, combining encryption and compression in a single step. A secret permutation p is used which is used to permute standard MPEG Huffman code word list. The permutation p must only permute the code words with the same number of bits to optimize compression ratio [8]. It is showed in [20] that Algorithm 1 is vulnerable to known-plaintext and cipher text-only attack. Knowing some of the video frames beforehand can enable the attacker to reconstruct the secret permutation p by comparing the known values with the encrypted frames [8]. According to [21], low frequency error attacks on Algorithm 1 define the cipher text-only attack. The permutation p only shuffles code words with the same length, so the most security comes from shuffling 16 bit code words in the AC coefficient table [8]. Since the number of code words with lengths less than 16 bits is limited, it is easy to reconstruct all of the DC coefficients and the most frequent AC coefficients as all of these will be encoded using less than 16 bit code words [8].

Algorithm 2 (VEA) was described in [18]. Only sign bits of the DC coefficients are encrypted in the I-frame blocks by XORing the sign bits with a secret key [8]. The length of key determines the level of security. There is a trade-off between length of the key and security. Too less key size is easier to break while very long key lengths are impractical while [8].

Shi et al. [19] developed Algorithm 4 (RVEA). It uses traditional symmetric key to encrypt the sign bite of the DCT coefficient and sign bite of the motion vectors. It is faster since only certain sign bites are encrypted in the MPEG stream [8]. This makes it better than the previous three techniques, in terms of security and is also 90% faster than Naïve approach [8].

The Selective Encryption Algorithm was developed in order to reduce processing overload for real time video applications [8]. Under this scheme, selective parts of the MPEG stream are encrypted. This based on the MPEG I-frame, P-frame and B-frame structure. Only I-frame is encrypted because conceptually, P-frame and B-frame are useless without the knowledge of the corresponding I-frame [8].

The AEGIS technique was developed by Maples and Spanos [23]-[24]. It only encrypts the I-frame of the MPEG stream. The sequence header is also encrypted which makes the MPEG video stream unrecognizable. The MPEG bit-stream is further hidden by encrypting the IOS end code. The DES encryption is used for the entire process. There is a tradeoff between video quality and security which means that, the quality degrades as the level of security is increased which depends on the length of the string [8].

## 3. IDEA OF THE PROPOSED SOLUTION

To fathom the algorithm, the process of encrypting multimedia is divided into two parts in this paper. The technique used for audio encryption is different from the technique used for the video while keeping the algorithm fast enough for real time compatibility. Both audio and video encryptions undergo two levels of encryption to ensure maximum safety. For video encryption, we use Pseudo Random Sequence Noise as level 1 and RSA algorithm for level 2, whereas, for audio encryption we use Pseudo Random Sequence Noise as level 1 and transform based encryption using discrete cosine transform. The reason for using these techniques and the algorithm will be explained in the subsequent sections.

## 4. VIDEO ENCRYPTION USING RSA ALGORITHM

RSA is a public-key cryptography algorithm, based on the presumed difficulty of factoring large integers, the factoring problem. RSA stands for Ron Rivest, Adi Shamir and Leonard Adleman, who first publicly described it in 1977 [25]. The RSA algorithm consists of three steps: key generation, encryption and decryption.

### 4.1 Key Generation

A key is a piece of information that determines the functional output of a cryptographic algorithm. Without a key, the algorithm would be useless. In encryption, a key specifies the particular transformation of plaintext into cipher text, or vice versa during decryption. There are two keys in RSA, i.e. Public key and Private key. The public key is known to everyone and is used for encrypting the messages; these messages can be decrypted only using the private key. Keys for the RSA algorithm are generated in the following manner:

(1) Select two distinct prime numbers *p* and *q*.
(2) Compute:
$$n = p \times q \qquad (1)$$
(3) Compute:
$$\varphi(n) = (p-1)(q-1) \qquad (2)$$
where $\varphi$ stands for the Euler's totient function
(4) Select an integer *e* such that $\varphi(n)$ and *e* are co prime.
(5) Calculate *d* using the formula
$$d \equiv e^{-1} \pmod{\varphi(n)} \qquad (3)$$

The public key consists of the modulus *n* and *e* (encryption exponent). The private key consists of the modulus *n* and *d* (decryption exponent), the decryption exponent has to be kept secret along with *p,q* and $\varphi(n)$, using which the decryption exponent can be calculated.

Let us consider an example for key generation:                (eg. 1)

(1) Let *p* and *q* be 73 & 89.
(2) $n = p \times q$ i.e. 73 × 89 which is equal to 6497.
(3) $\varphi(n) = 72 \times 88$ which is equal to 6336.
(4) We select an integer 113 such that $\varphi(n)$ and 113 are co-prime.
(5) *d* is calculated using the formula; *d* = 785.

### 4.2 Encryption

Encryption is the process of transforming information using an algorithm to make it unreadable to anyone except those who possess the key to decrypt it. Here's an example illustrating how plaintext is encrypted using the keys created in the e.g. 1 of key generation.

Plaintext message: ravsushaman                (eg. 2)
Plaintext Message in numeric form: 114 097 118 115 117 115 104 097 109 097 110

$$c = m^e \pmod{n} \qquad (4)$$

Let us encrypt the first part of the message, 114, the calculation would be c = 114113(mod 6497) which is equal to 6369, similarly the entire the block of message is converted.

Encrypted message in numeric form: 6369 6208 3903 3077 3040 3077 5756 6208 3926 6208 1330

#### 4.2.1 Steps of encryption
(1) The AVI file is loaded into the system.
(2) The frames of the file loaded, is extracted one by one.

(3) After the extraction process, the frames are loaded.
(4) The loaded frames are then segregated into their RGB components and the encryption takes place on the individual color components of the frame.
(5) The RGB frames are encrypted individually using the RSA algorithm.
(6) To initiate the RSA process, accept an input string from the user. The sum of the ASCII values of each character of the string input by the user is stored as *x*. Two large consecutive prime numbers are selected which are immediately next to *x* and pass them on as inputs to the RSA algorithm.
(7) Key distribution takes place.
   (a) The public key is sent from the sender to the receiver.
   (b) The receiver then sends a message *M* to the sender.
   (c) The message *M* is first converted into an integer m, such that $0 < m < n$ by using an agreed-upon reversible protocol known as a padding scheme.
   (d) The cipher text c is calculated corresponding to $c = m^e (mod\ n)$.
   (e) The cipher text *c* is then sent from the receiver to the sender.
(8) The encrypted RGB components are then combined as a JPG file (with a frame number in the filename).
(9) Steps 3 to 6 for all frames are repeated for all the frames.
(10) The result is obtained after utilizing all the stored frames to create a video file with each stored encrypted image as an individual frame of the video.

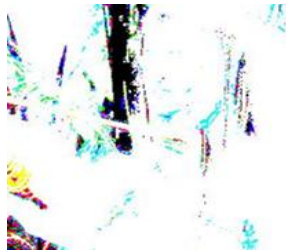

**Fig 2: Video after encryption.**

## 4.3 Decryption
Decryption is the process of extracting the original message from a ciphertext using the algorithm which requires a key to decrypt the message.

$$m = c^d\ (mod\ n) \qquad (5)$$

We decrypt the message *m* using the above formula.

Here is an example for decryption using the key and cipher from e.g. 1 & 2:  (eg. 3)

Let us decrypt the first part of the encrypted message, 6369, using the formula, $m = 6369^{785}(mod\ 6497) = 114$. Similarly, we repeat the process for the whole text message.

Decrypted message in numeric form: 114 097 118 115 117 115 104 097 109 097 110

### 4.3.1 Steps of decryption
(1) The encrypted AVI video file is loaded into the system.
(2) The frames are extracted one by one.
(3) Each frame is loaded in the system.
(4) The RGB components are then extracted from the loaded frames.
(5) The RGB components are decrypted using RSA.
(6) The sender recovers *m* from *c* by using the private key exponent via computing $m = c^d\ (mod\ n)$.
(7) Each decrypted RGB component is then combined as a JPG file (with a frame number in the filename).
(8) Steps 3 to 6 are repeated for all frames.
(9) The result is obtained after utilizing all the stored frames to create a video file with each stored decrypted image as an individual frame of the video.

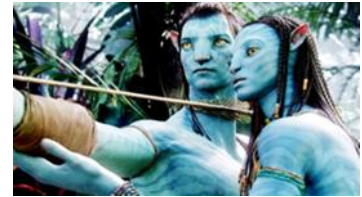

**Fig 3: Video after decryption.**

## 5. VIDEO ENCRYPTION USING RSA ALGORITHM AND PSEUDO RANDOM SEQUENCE NOISE
We introduce another level of encryption in this section, which uses pseudo random sequence noise. The need for two levels is explained in the subsequent sections.

### 5.1 Pseudorandom generators
A pseudorandom generator [26] (PRG) is a deterministic map $G:\{0,1\}^l \to \{0,1\}^n$, where $n \geq l$. Here, *n* is the "seed length" and *l* is the "stretch". Usually $n \gg l$ and G is efficiently computable in some model. If $f:\{0,1\}^n \to \{0,1\}$ is any "statistical test", we say G "$\varepsilon$-fools" *f* is

$$\left|\Pr[f(U_n) = 1] - \Pr[f(G(U_l)) = 1]\right| \leq \varepsilon \qquad (6)$$

Where $U_m$ denotes an uniformly random string in $\{0,1\}^m$. Here the string $U_i$ is called the "seed". If *C* is a class of tests, we say that G "$\varepsilon$-fools C" or is an "$\varepsilon$-PRG against C" if G $\varepsilon$-fools *f* for every $f \in C$

#### 5.1.1 Alternate definition
A deterministic function $G:\{0,1\}^d \to \{0,1\}^m$ is *(l,ε)* pseudorandom generator (PRG) if,
(1) $d < m$, and
(2) $G(U_d)$ and $U_m$ are *(t,ε)* indistinguishable

### 5.2 Pseudorandom functions
Pseudorandom functions [26] are like pseudo random generators whose output is exponentially long and it is such that given a seed, each bit of the output is computable. The security is against efficient adversaries that are allowed to look at any subset of the exponentially many output bits.

#### 5.2.1 Definition
A function $F: \{0,1\}^k \times \{0,1\}^m \to \{0,1\}^m$ is a *(t,ε)* secure pseudorandom function if for every oracle algorithm *T* that has complexity at most *t* we have [26],

$$\left| P_{K\in\{0,1\}^k}[T^{F_k}() = 1] - P_{R:\{0,1\}^m \to \{0,1\}^m}[T^R() = 1] \right| \leq \varepsilon \qquad (7)$$

This implies that it is impossible to distinguish outputs from a pseudorandom function (up to a certain additive *ε* error) and a purely random function. Typical parameters are $k = m = 128$. This gives a security high of $(2^{60}, 2^{-40})$ [26].

## 5.3 Encryption using pseudorandom functions

Suppose $F : \{0,1\}^k \times \{0,1\}^m \to \{0,1\}^m$ is a pseudorandom function. The encryption scheme is defined as [26]:

- $Enc(K, M)$ : pick a random $r \in \{0,1\}^m$, output $(r, F_K(r) \oplus M)$

- $Dec(K,(C_0, C_1)) := F_K(C_0) \oplus C_1$

## 5.4 Need for Two Levels of Encryption

Upon implementing RSA encryption for the video, the encrypted video turned out to visually resemble the original video, we can see it clearly in Fig 4 below that the encrypted image bears some similarities to the original image. For highly sensitive data it is imperative that the encrypted video should not bear any visual resemblance to the original video. Hence, we try to remove any resemblance whatsoever from the encrypted video to the original video, to do this we propose the use of two levels of encryption.

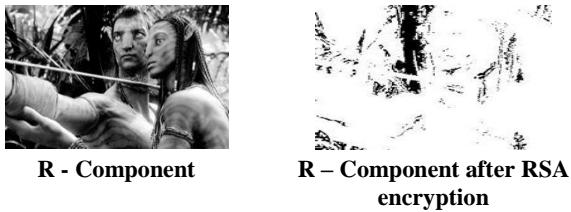

**R - Component**     **R – Component after RSA encryption**

**Fig 4: R – Component after RSA encryption**

In our proposed system the first level encryption is obtained by Pseudo Noise sequence whereas, level 2 encryption is obtained by using RSA algorithm. The result of combining these two levels is shown in Fig 5 below.

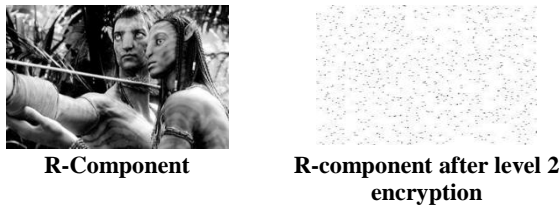

**R-Component**     **R-component after level 2 encryption**

**Fig 5: Comparison of image before and after 2 levels of encryption**

From Fig. 5 above we can see that there is no resemblance between the original and encrypted image, which is essential to provide maximum security. The entire process is depicted in Fig 6 (refer last page of article).

## 6. AUDIO ENCRYPTION

It is relatively easier to crack single level encryption by brute force or correlation, as compared to a multi-level encryption scheme. Therefore, the program requires a user defined 8-character password to seed two out of three levels of encryption. The password must have a minimum of one capital alphabet, one numeral and one special character, such as @, >, & etc., leading to an excess of $6.6 \times 10^{15}$ possible permutations. This password is processed further using numeric substitutions in Caesar's Cipher type of encryption and a state seed is generated. Findings [5] show that a hacker may take as less as 10 minutes to crack each password once a rainbow table has been built, if all passwords are stored internally in a memory hash. Hence, the system is designed to store no password and the state seed generated on the fly is divided into two keys, each of which is used for further encryption and computation.

We adopt the audio encryption scheme as outlined in [27], which can be summarized as following:

- A Pseudo Random Number Generator (PNRG) can be used to generate a statistically independent and unbiased stream of binary digits.
- The PNRG is implemented via the inbuilt MATLAB 7 function, in which a random sequence on average, will repeat only after $235 \times 16$ bits. This is greater than the length of samples required for testing, making it difficult for an adversary to figure out the next bit by correlation of the subset of random bits generated [12].
- A disadvantage of such a technique is that the signal is still in time domain, and deciphering the signal is possible by guessing the scrambling key.
- A popular approach to this is the Discrete Cosine Transform (DCT), which transforms the time domain signal to the frequency domain signal as shown in [11].
- DCT coefficients give the frequency domain equivalent of the analog speech signal [11] by decomposing the signal into its frequency components and un-correlating the sequence of input samples.
- Such a technique can be used in large databases, for voice signal compression, throwing away some of the accuracy.
- A layer of pseudorandom noise is added on top of the DCT coefficients, using the second part of the state key as an added layer of protection.
- This signal is amplified and transmitted to the receiver side over an AWGN channel with a known Signal to Noise Ratio (SNR), where it is decrypted by following the reverse order of the encryption process.

### 6.1 Implementation Steps

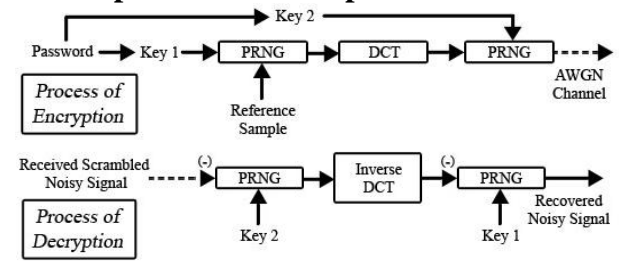

**Fig 7: Scheme for encryption and decryption**

The steps involved in the process of generation of the two State Keys, which have been outlined in [27], can be summarized as follows:

(1) The user provides an 8 character long password.
(2) An ASCII equivalent of the password is formed as an array denoted by $x$.
(3) A new array, $y$ is created by applying Caesar's cipher with a shift of 4 to each ASCII element.
(4) A single integer, $z$ is formed by concatenating all elements of $y$.
(5) Length of $z$ is calculated.

(6) If the length is even, it is divided into two equal halves to form two equal length keys – Key 1 and Key 2.
(7) If the length is uneven, it is divided asymmetrically, the longer key being Key 1 and shorter key being Key 2.

The steps involved in the process of encryption can be summarized as follows:

(1) Level 1: A noisy signal $x$ is generated by passing Key 1 as a seed to PNRG and superimposing the output on the reference voice sample.
(2) Level 2: A new array $y$ is formed by performing DCT on $x$ and saving the coefficients.
(3) Level 3: A new array $z$ is formed by passing Key 2 as a seed to PNRG and superimposing the obtained random sequence on $y$.
(4) The signal which is transmitted through AWGN channels $z$, has known SNR. Thus, it is subjected to various SNR levels and the resulting signal is then normalized.

The steps involved in the process of decryption can be summarized as follows:

(1) Level 3: Key 2 is passed as seed to PRNG and the random sequence generated is algebraically subtracted from the received signal, saved as $x$.
(2) Level 2: The noisy signal $y$ is generated by performing Inverse DCT on $x$.
(3) Level 1: Key 1 is passed as seed to PRNG to generate a random sequence which is algebraically subtracted from $x$ and saved as $z$, which is the final decrypted version of reference signal.

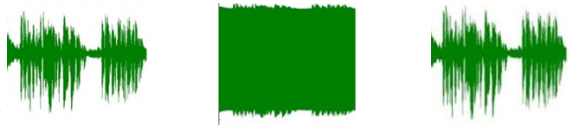

**Fig 8: Original signal (Left), signal after Level 2 encryption (Middle) and Recovered file with high SNR (Right)**

## 7. FACTORS AFFECTING THE PERFORMANCE OF ENCRYPTION

**Encryption/Decryption Speed**: This criterion is the measure of the time taken to encrypt/decrypt a color frame of the video. The time taken should be minimum to increase the efficiency of the system.

**Cryptographic Security (CS):** Cryptographic security defines whether encryption algorithm is secure against brute force and different plaintext-ciphertext attack. For highly valuable multimedia application, it is really important that the encryption algorithm should satisfy cryptographic security.

$$CS = \frac{Total\ time\ taken}{Number\ of\ frames\ in\ the\ video} \quad (8)$$

**Encryption Ratio (ER):** This criterion measures the ratio between the size of encrypted part and the whole data size. Encryption ratio has to be minimized to reduce computational complexity.

$$ER = \frac{Original\ size}{Encrypted\ size} \times 100 \quad (9)$$

**Decryption Ratio (DR):** This criterion measures the ratio between the size of encrypted part and the whole data size. Encryption ratio has to be minimized to reduce computational complexity.

$$DR = \frac{Original\ size}{Decrypted\ size} \times 100 \quad (10)$$

**Visual Degradation (VD):** This criterion measures the perceptual distortion of the video data with respect to the plain video. In some applications, it could be desirable to achieve enough visual degradation, so that an attacker would still understand the content but prefer to pay to access the unencrypted content. However, for sensitive data, high visual degradation could be desirable to completely disguise the visual content.

To test our system we use a short clip of three seconds of size 244 KB, Table 1 shows the details of the results for encrypting and decrypting the video. We measure the performance of these algorithms on an Intel Core 2 Duo 2.6 GHz machine with 2 GB of RAM.

**Table 1. Sample results**

| Quality | Value (with PN & RSA) | Value (with only RSA) |
|---|---|---|
| Encrypted Size | 1.54 MB | 477 KB |
| Decrypted Size | 677 KB | 677 KB |
| Encryption Time (90 frames) | 204.3 s | 190.16 s |
| Decryption Time (90 frames) | 287.1 s | 220.17 s |

Table 2 shows the values of the parameters on which the performance of our system depends on.

**Table 2. Parameter Details**

| Quality | Value (with PN & RSA) | Value (with only RSA) |
|---|---|---|
| Encryption Ratio | 16% | 50.7% |
| Decryption Ratio | 36% | 36% |
| Visual Degradation | No visual resemblance | Slight resemblance |
| Speed (Encryption per color frame) | 2.27 s | 2.08 s |
| Speed (Decryption per color frame) | 3.19 s | 2.49 s |

## 8. CONCLUSION

We have proposed a new method of dual layer encryption methodology which enables to achieve zero visual resemblance and high security while not being severely penalized in Speed and Decryption ratio. We achieved an Encryption ratio of 16% using PN and RSA technique as compared to 50.7% using just the RSA technique. The Decryption is the same for both approaches at 36%. The highlight of this approach is low penalty in Encryption and Decryption speed. The dual layer approach took just 0.19 seconds/frame and 0.7 seconds/frame more for Encryption and Decryption respectively as compared to the RSA only approach. So, there is less than 1 second penalty involved while achieving zero visual resemblance and respectable Encryption and Decryption ratios.  This is achieved by separating the audio and video frames and applying the techniques individually rather than encrypting and decrypting all the frames in one go. Potential speed improvements in performance could be possible by incorporating a technique similar to the selective encryption algorithm [8], discussed before. Another possible scheme can involve the two step audio encryption process be selectively relaxed based on the content of the media.  Such a technique can be very beneficial to on-demand audio-visual entertainment services like Netflix, etc. The dual layer approach presents a promising approach to achieving a highly secure way of video encryption while not being very computationally intensive and time consuming.

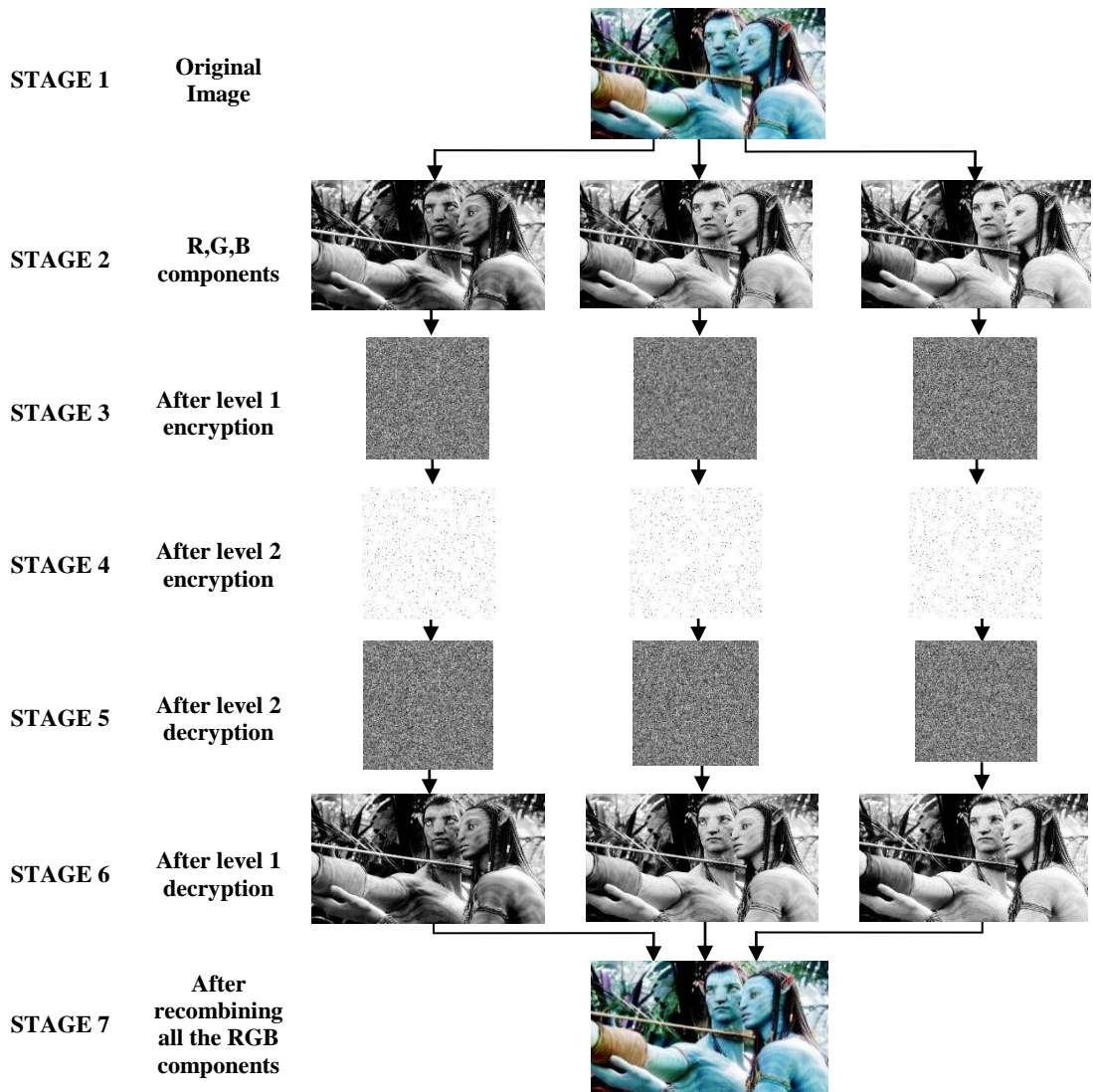

**Fig 6: Encrypting and decrypting over the RGB Components**

## 10. AUTHOR'S PROFILES

**Aman Chadha** (M'2008) was born in Mumbai (M.H.) in India on November 22, 1990. He completed his M.S. in Electrical and Computer Engineering (with a specialization in Computer Architecture) at the University of Wisconsin-Madison, where he was conferred the Outstanding International Graduate Student Award in 2014. He completed his B.E. in Electronics and Telecommunication Engineering with Distinction from the University of Mumbai in 2012. His fields of interest include Design of Specialized Computer Architectural Systems, Signal and Image Processing, Computer Vision (particularly, Pattern Recognition) and Processor Microarchitecture. Along with 12 papers in International Conferences and Journals and 2 books to his credit, he has been conferred numerous awards and scholarships at the graduate and undergraduate level. He is a member of the IETE, IACSIT, ISTE and ACM.

**Sushmit Mallik** (M'2008) was born in Kolkata (W.B.) in India on October 12, 1990. He completed his M.S. in Electrical Engineering (with a specialization in Nano-electronics and Photonics) at North Carolina State University. He completed his B.Tech. in Electronics and Communication Engineering from SRM University in 2012. Previously, he was a visiting student at the University of Wisconsin-Madison in 2011 and a Research Assistant at the University of Hong Kong in 2012. He has 3 papers in International Conferences and Journals.

**Ankit Chadha** (M'2008) was born in Mumbai (M.H.) in India on November 7, 1992. He is pursuing his M.S. in Electrical and Computer Engineering (with a specialization in VLSI) at the University of Minnesota. He completed his B.E. in Electronics and Telecommunication Engineering with Distinction from the University of Mumbai in 2012. He has 8 papers in International Conferences and Journals.

**Ravdeep Johar** (M'2008) was born in Bokaro (J.H.) in India on July 16, 1991. He completed his M.S. in Computer Sciences at the Rochester Institute of Technology and his B.Tech. in Electronics and Communication Engineering from SRM University in 2012. Previously, he was a visiting student at the University of Wisconsin-Milwaukee in 2011.

**M. Mani Roja** (M'1990) was born in Tirunelveli (T.N.) in India on June 19, 1969. She completed her B.E. in Electronics and Communication Engineering from GCE Tirunelveli in 1990 and M.E. in Electronics from Mumbai University in 2002. Her employment experience includes 24 years as an Associate Professor at Thadomal Shahani Engineering College (TSEC), Mumbai University. She has over 20 papers in National/International Conferences and Journals to her credit. She is a member of IETE, IACSIT, ISTE and ACM.